# Filamentary condensations in a young cluster

**Philip C. Myers**

Harvard-Smithsonian Center for Astrophysics, 60 Garden Street, Cambridge

MA 02138 USA

pmyers@cfa.harvard.edu

**Abstract.** New models are presented for star-forming condensations in clusters. In each model, the condensation mass increases linearly with radius on small scales, and more rapidly on large scales, as in "thermal-nonthermal" models. Spherical condensations with this structure form protostars which match the IMF if their infall is subject to equally likely stopping. However such spherical models do not match the filamentary nature of cluster gas, and they are too extended to form protostars having high mass and short spacing. Two hybrid models are presented, which are spherical on small scales and filamentary on large scales. In and around clusters, cores embedded in linear filaments match the elongation of cluster gas, and the central concentration of low-mass stars. In cluster centers, condensations require a low volume filling factor to produce massive stars with short spacing. These may have stellate shape, where cores are nodes of filamentary networks, as seen in some simulations of colliding flows and of collapsing turbulent clumps. A dense configuration of such stellate condensations may be indistinguishable from a clump forming multiple protostars via filamentary flow paths.

# 1. Introduction

How star clusters form is a long-standing problem in astrophysics (Clarke 2010, Lada 2010). It is still not understood how hundreds of stars form in centrally condensed configurations of ~ 1 pc extent over ~1 Myr, having a mass distribution following the initial mass function (IMF). The problem of forming such IMF-clusters is further constrained by recent observations indicating that much of the dense gas in young clusters is organized into filaments with embedded cores, having peak column densities exceeding $10^{22}$ cm$^{-2}$ (Allen et al 2007, Gutermuth et al 2009, Myers 2009a).

Termination of star-forming infall is expected to be a key process in setting protostar mass, and thus in setting the IMF (Shu, Adams & Lizano 1987). Yet it remains unclear how infall is terminated. In "monolithic collapse" models, the typical infall is assumed to endure until a fixed fraction of the gas within a core boundary has fallen onto the protostar-disk system (e.g. McKee & Tan 2003, Alves, Lada & Lada 2007). However, the physical basis of this gas fraction and of the core boundary are not well understood. In hydrodynamic simulations of competitive accretion, accretion is limited by the gravitational dynamics of ejection and of multiple accretors (e.g. Bonnell et al 1997), but gas-pressure forces due to outflows and stellar feedback are usually not included. Further, computational limitations make it difficult to follow a significant number of protostars until they achieve their final masses.

Although it is uncertain how infall is terminated, the duration of infall can be described statistically. In a young cluster, the combination of dynamical ejection, dispersal of dense gas by stellar feedback, competitive accretion and other processes has been proposed to terminate infall with equal likelihood at any moment (Myers 2009). If so, the probability density of infall durations is a declining exponential function of time, with a single parameter, the mean infall duration (Myers 2000, Basu & Jones 2004).

When the distribution of infall durations is specified, it is possible to predict the protostar mass function arising from the cold spherical collapse of an initial core-environment configuration. It was found that cores in environments with diverse geometrical structure can produce mass functions which are good approximations to the IMF, including its high-mass tail, provided the environment gas is dense enough, typically of order $10^4$ cm$^{-3}$ (Myers 2009).

Conversely, this same distribution of durations was used to find the spherically symmetric density profile whose collapse generates the IMF exactly, over a protostar mass range of 0.05 - 10 M$_\odot$ (Myers 2010, hereafter Paper 1).

This density profile which generates the IMF has a steep-slope component resembling the singular isothermal sphere model of a dense core (SIS; Chandrasekhar 1939, Shu 1977), and a shallow-slope component resembling the "clump" gas surrounding a dense core, as observed in nearby star-forming regions (Teixeira, Lada & Alves 2005, Kirk, Johnstone & Di Francesco 2006). This profile resembles the two-component thermal and nonthermal (TNT) model of Myers & Fuller (1992) and the two-component turbulent core (2CTC) model of McKee & Tan (2003). The core and clump components are each necessary, to account for both the low-mass peak and high-mass tail of the IMF.

This core-clump model provides a simple way to understand the origin of the IMF, with very few assumptions. However, it remains to explain how these condensation properties relate to the spatial concentration of protostars in a cluster, and to the filamentary structure of the protocluster gas.

The stars in young embedded clusters are centrally concentrated, following the structure of their parent molecular gas (Allen et al 2007). They lack the spherical symmetry of more evolved clusters, and their surface density maps often have more than one peak. Nonetheless the spacing of their young stars is distinctly closer than for surrounding stars, with typical nearest-neighbor projected spacing 0.07 pc in the central regions of 36 nearby young clusters (Gutermuth et al 2009).

The gas associated with such young clusters has "filamentary" structure, indicated by linear features in maps of dust extinction at optical and near infrared wavelengths, of dust emission at submillimeter wavelengths, and of molecular line emission (Wiseman & Ho 1998, Johnstone & Bally 1999, Motte et al 2001, Nutter et al 2006, Enoch et al 2006, Myers 2009a). Recent observations by the *Spitzer Space Telescope* and the *Herschel Space Observatory* have revealed more such structure, suggesting it is common in young clusters and their associated molecular clouds. The Serpens South region in Aquila hosts a spatially concentrated cluster with a high fraction of protostars, embedded in a region of high column density, with numerous dense gas filaments radiating outward (Gutermuth et al 2008, André et al 2010, Molinari et al 2010).

This paper extends the model of Paper 1 to describe these features of cluster structure. Section 2 describes condensation models whose geometry is purely spherical, or is spherical on small scales and filamentary on large scales. Section 3 discusses the results, and Section 4 summarizes the paper.

## 2. Condensation models

This section describes three star-forming condensations having the same dependence of mass on radius. This dependence is chosen to allow an ensemble of such condensations to form protostars whose mass function approximates the IMF, in a simple model of cold spherical infall with equally likely stopping. The "spherical" condensation has spherical symmetry, and is similar to that in Paper 1. The "linear filament" condensation has a spherical core in a cylindrical clump, and is part of the linear filamentary chains which form mainly low-mass stars. The "hub-filament" condensation has a spherical core which radiates multiple dense filaments, in the central regions of clusters which form massive stars.

As condensation shape departs from spherical symmetry, the model of cold spherical infall become less accurate in predicting the mass of the protostar at a given infall duration. Since each model considered here has the same spherically symmetric core, all three models will have nearly identical relations of mass to free-fall time for small radii, short infall duration, and low mass. Therefore their protostar mass functions can be expected to be very similar for low-mass stars. They are likely to differ more in their large-scale infall motions, due to nonradial gravitational forces and fragmentation, and the corresponding mass functions are likely to differ more in their high-mass tails. Numerical calculations are required to quantify this difference, but such calculations are beyond the scope of this paper. This point is discussed further in Section 3.2.

Thus the requirement that each condensation have the same dependence of mass on radius is useful to constrain condensation models, and its resulting protostar mass functions are likely to be similar, but not identical.

2.1. Spherical condensation

Paper 1 showed that an ensemble of spherically symmetric condensations having density profiles with a steep-slope core and a shallow-slope clump can be combined with a simple collapse model subject to equally likely stopping. The resulting distribution of protostar masses matches the IMF, provided the density $n_s$ has the form

$$n_s = \frac{A}{r^2} + \frac{B}{r^{2/3}} \qquad (1)$$

where $r$ is the spherical radius. Then integration gives the mass $m_s$ within $r$ as

$$m_s = 4\pi m A r + \frac{12\pi m B r^{7/3}}{7} \qquad (2)$$

where $m$ is the mean molecular mass.

The core component has the structure of a singular isothermal sphere (SIS; Chandrasekhar 1939) with temperature $T$, so that the coefficient $A$ is given by

$$A = \frac{kT}{2\pi m^2 G} \quad . \qquad (3)$$

Here $k$ is Boltzmann's constant and $G$ is the gravitational constant. The spherical clump component can be expressed in terms of the maximum radius $R$ and the mean column density $\overline{N}_s$ within this projected radius, according to

$$B = \frac{7\overline{N}_s}{12R^{1/3}} \qquad (4)$$

so that $m_s$, the spherical condensation mass within radius $r$, is given by

$$m_s = \frac{2kTr}{mG} + \frac{\pi m \overline{N}_s r^{7/3}}{R^{1/3}} \qquad (5)$$

from equations (2), (3), and (4).

The density profile corresponding to equation (1) is shown in Figure 4 in Paper 1. The dependence of mass on radius in equation (5) is shown in Figure 1 of this paper, for $T = 15$ K, $\overline{N}_s = 10^{22}$ cm$^{-2}$, and $R = 0.5$ pc. There, the increase of mass with radius is linear for condensation gas dominated by the core component, and is steeper than linear for condensation gas dominated by the clump component. The mass is 1.9 $M_\odot$ within 0.05 pc and 158 $M_\odot$ within 0.5 pc.

In Paper 1 a limiting condensation radius and mass were assigned at the radius where the gradient of mean column density with radius becomes flat. Here similar values are assigned, but with a more physical basis.

In nearby star-forming regions, gravitational competition from neighboring condensations probably limits available mass to maximum radii of a few 0.1 pc. The projected spacing of condensations is greatest in regions of isolated star formation, such as the Pipe Nebula, with spacing 0.4 pc (Rathborne et al 2009), and in Taurus, with mean spacing 0.3 pc (Hartmann 2002). In the densest part of young clusters such as NGC 1333, this spacing decreases to 0.1 pc (Rosolowsky et al 2008). The mass available for infall is also limited when a protostar is dynamically ejected from a multiple system in dense gas to a less dense medium (Reipurth & Clarke 2001, Bate 2005). Similarly, gas pressure forces can remove or redirect low-density gas before it can fall into a center of gravitational attraction. Pressure forces from stellar feedback can limit infall due to ionization, heating, shocks, winds and outflows, as discussed in Paper 1. The effect of such dispersal is evident over several 0.1 pc in NGC 1333, where some regions have relatively low extinction and harbor mainly T Tauri stars, indicating removal of dense gas during the T Tauri lifetime of ~ 1 Myr (Gutermuth et al 2008).

2.2. Linear filament condensations

To represent observed filamentary clouds more realistically than in the spherically symmetric condensation model, a "linear filament" condensation is described, where the clump symmetry is cylindrical rather than spherical. The density is given by

$$n_f = \frac{A}{r^2} + \frac{C}{b^{2/3}} \qquad (6)$$

where $b$ is the cylindrical radius and $C$ is a constant. The clump component is axisymmetric about the $z$ axis, so that

$$b^2 = x^2 + y^2 \qquad (7)$$

and

$$r^2 = b^2 + z^2 \quad . \tag{8}$$

The density dependence of the cylindrical clump component, as $b^{-2/3}$ in equation (6), is chosen to match the density dependence of the spherical clump component, as $r^{-2/3}$ in equation (2). For any such power-law exponent, the mass within a given spherical radius can have the same dependence on radius in both spherical and cylindrical geometry, as is shown below.

Like the spherical condensation, each linear filament condensation has a maximum mass available for formation of a single star, due to gravitational competition from neighboring condensations and to gas dispersal by stellar feedback. The corresponding limiting distances $|z_{max}|$ along a filament axis, and $b_{max}$ across the axis, are represented for simplicity by the same constant parameter,

$$|z|_{max} = b_{max} = s/2 \quad , \tag{9}$$

where $s$ is the typical spacing of neighboring cores along a filament axis. Observations suggest that a reasonable choice of $s$ is 0.2 - 0.3 pc as discussed in Section 2.1 above. The limiting filamentary condensation radius $s/2$ is analogous to the limiting spherical condensation radius $R$. In a more realistic model, $b_{max}$ should be allowed to differ from $|z_{max}|$, and $s$ should be allowed to decrease from the outer to the inner parts of a protocluster. In the hub-filament model of Section 2.5, $z_{max} \gg b_{max}$ for each of several filaments associated with a core.

In equation (6) the coefficient $A$ is the same as is given in equation (3), since the spherical and filamentary condensations have the same core component. The coefficient $C$ is expressed in terms of $s/2$ and $\overline{N}_f$, the mean filamentary clump column density within projected radius $s/2$, according to

$$C = \frac{\overline{N}_f}{3 I_1 (s/2)^{1/3}} \tag{10}$$

where $I_1$ is a constant due to integration of the density profile in equation (6) along the z-direction,

$$I_1 \equiv \int_0^1 d\zeta (1-\zeta^2)^{2/3} = 0.739, \tag{11}$$

where the dummy variable $\zeta$ is defined by $\zeta \equiv z/r$. Then the condensation mass within spherical radius $r$ is obtained for the linear filament by integration of the density profile in equation (6) as

$$m_f = \frac{2kTr}{mG} + \frac{\pi m \overline{N}_f r^{7/3}}{(s/2)^{1/3}} \tag{12}$$

for $r \leq s/2$.

Comparison of equations (5) and (12) shows that the linear filament and spherical condensations have identical dependence of enclosed mass on spherical radius, if they have equal maximum axial and radial sizes $b_{max} = |z_{max}| = R$, equal mean column densities $\overline{N}_s = \overline{N}_f$ within this projected radius, and the same core temperature $T$. This equality is exact, but it does not apply to the relatively small amount of mass in the filamentary clump beyond radius $s/2$, in the "corners" between spherical radii $b_{max}$ and $(b_{max}^2 + z_{max}^2)^{1/2}$.

Typical values of core, clump, and condensation masses are estimated from observations of the nearest and youngest cluster-forming regions. The typical maximum radius is 0.15 pc based on condensation spacings as discussed above. The mean clump column density is $1 \times 10^{22}$ cm$^{-2}$ as in nearby star-forming clouds (Motte, Andre & Neri 1998, Teixeira et al 2007, Gutermuth et al 2008). The kinetic temperature is 15 K, based on NH$_3$ lines in clouds associated with young clusters (Jijina, Myers & Adams 1999). Then equations (5) and (12) give $m_s = m_f = 17 \ M_\odot$, with 3.7 $M_\odot$ in the core component and 13 $M_\odot$ in the clump component.

With these parameters the mass in the filamentary core-clump condensation, including the corners, exceeds that within the spherical radius 0.15 pc by a factor 1.23. This increase applies only to the largest scales and to the most massive stars formed by such condensations. This increase would not be noticeable in mass function histograms having bin widths of a factor of two.

Similarly, a crude estimate of the mass accretion rate is the total mass divided by the free-fall time corresponding to the mean density. This mean mass accretion rate for the entire condensation is 88 $M_\odot$ Myr$^{-1}$, and it exceeds that of its spherical component alone by a factor 1.18. For the purposes of the present paper, this relative increase in mean mass accretion rate is negligibly different from unity.

2.3. Chain of filamentary condensations

Filamentary dark clouds are generally "globular" with embedded condensations, and in some regions these condensations have similar spacing (Schneider & Elmegreen 1979, Curry 2000). Such filamentary chains are also seen in cluster-forming regions such as NGC 1333 (Sandell & Knee 2001, Walsh et al 2007). Observations with the *Herschel Space Observatory* indicate that filaments with multiple condensations are common in the Serpens South cloud and in other parts of the Aquila star-forming complex (André et al 2010, Molinari et al 2010).

To represent the structure observed in filamentary star-forming clouds, the single condensation described in section 2.2 above is assumed to be one of many identical condensations joined in filamentary chains. Neighboring condensations are assumed to be coaxial and to join at $|z| = s/2$.

Figure 2 shows contours of constant column density for linear filament condensations with the parameters $T = 15$ K, $\overline{N} = 1 \times 10^{22}$ cm$^{-2}$, and $s = 0.3$ pc assumed above. The filamentary condensation resembles observed clouds since it is elongated and filamentary on large scales, but is more nearly round on small scales.

The chains of filamentary star-forming condensations in Figure 2 resemble observed filaments in their shape, extinction, and in their association with young low-mass stars. Such filamentary chains may provide a central concentration of birthsites for low-mass stars, if the chains converge toward one or more central positions. Converging filaments are observed on the parsec scale in several young clusters and infrared dark clouds (Myers 2009, André et al 2010), and converging filaments are seen in some simulations of cluster-forming regions (Vazquez-Semadeni et al 2007, Banerjee et al 2010).

Converging filamentary chains can provide some of the central concentration of protostars seen in young clusters. For $N_{fil}$ radial filaments in the plane of the sky, having condensations with equal spacing $s$, the mean surface density of protostar birthsites is $N_{fil}/(\pi s r)$ for $r > s$, increasing inward as $r^{-1}$. Then a system of 10 converging filaments of radius 0.4 pc, having condensations with spacing 0.2 pc, has a mean surface density of birthsites 40 pc$^{-2}$, close to the typical mean surface density of protostars, 60 pc$^{-2}$, in a sample of 36 young clusters within 1 kpc of the Sun (Gutermuth et al 2009).

The filamentary condensations considered here are lumpier and have greater mass per unit length than do the smooth filaments in models based on isothermal equilibrium, or on isothermal and magnetostatic equilibrium (Stodolkiewicz 1963, Ostriker 1964). In these models, when the radial distance $b$ from the symmetry axis is much greater than the scale length, the density falls off as $b^{-4}$. This steep decline may also apply to radially collapsing filaments (Tilley & Pudritz 2003). In contrast, in the present filamentary condensation models, the density falls off as $b^{-2}$ where the core component dominates and as $b^{-2/3}$ between neighboring cores, where the clump component dominates. As new observations of dust emission become available, it should be possible to measure the structure of star-forming filaments much more accurately than has been possible previously. A study based on dust extinction found that density declines approximately as $b^{-2}$ in the filamentary cloud L977 (Alves et al 1998).

2.4. Spacing of massive star condensations

The foregoing spherical and linear filament condensations are space-filling structures whose ratio of shortest and longest dimensions is unity. These properties imply that these condensations cannot simultaneously account for the masses of massive stars, and for the close spacings of stars in young clusters.

The typical spacing of young stars near cluster centers in 36 embedded clusters within 1 kpc of the Sun is 0.07 pc (Gutermuth et al 2009), and in more massive clusters the separations of massive stars from their nearest neighbors is comparable to, or less than this value. In the central 0.1 pc of the Orion Nebula Cluster the typical spacing of cluster members is closer to 0.01 pc, and there is no obvious dependence of the nearest-neighbor spacing on the mass of one of its members (Hillenbrand 1997, Figure 3). Such close spacing could arise as an artifact of projection, but only in the highly unlikely case that the distribution of stars is much more extended along than across the line of sight.

The close spacing of stars in young clusters is probably not due to initially larger spacings, followed by dynamical concentration. It is expected that massive stars will decrease their spacing as they sink toward the center of a cluster through two-body interactions, giving kinetic energy to lower-mass stars. However this process takes many cluster crossing times, and cannot explain the observed mass segregation in young clusters such as as the Trapezium in Orion (Bonnell & Davies 1998). The same process would tend to gradually increase the spacing of lower-mass stars, not to decrease them.

If the initial configuration of stars is sufficiently clumpy and subvirial, the degree of mass segregation can increase dynamically in ~ 1 Myr, as clumps of stars fall toward their common center of gravity, according to a numerical simulation (Allison et al 2009). However this process requires the initial star clumps to be dense, well-separated, subvirial, and free of gas. These conditions may not be realistic enough to apply to centrally concentrated, gas-rich young clusters.

In contrast to the proposed concentration of cluster stars with time, observations of young clusters indicate that cluster cores of similar mass follow a trend of increasing core radius with age, over ages 1-20 Myr. This trend may be due to a combination of gas expulsion, stellar mass loss, and heating by stellar-mass black holes (Bastian et al 2008).

To make a star of mass 10 $M_\odot$ with accretion efficiency 0.5 as discussed in Section 2.1 above, a condensation having IMF-matching density profiles as in Sections 2.1 and 2.2 must have diameter 0.3 pc in each dimension. Thus the birthsites of such massive stars should be separated from those of their nearest neighbor stars by at least 0.3 pc, even in the limiting case where the condensations are close-packed and where their neighbor stars have smaller initial condensations than the massive stars. This initial separation is much greater than the typical protostar spacing near cluster centers.

In contrast, formation of low-mass stars with the present model does not face such a clear spacing problem. To match the peak of the IMF by forming a protostar of mass 0.16 $M_\odot$ with accretion efficiency 0.5, the spherically symmetric model of Section 2.1 would require a condensation of initial radius 0.01 pc, comparable to the typical spacing in the central part of the Orion Nebula Cluster and much less than the typical spacing in nearby embedded clusters. A spacing problem would apply to these low-mass stars only if initial condensations were required to have the same extent for protostars of all masses.

This spacing problem is not limited to the particular condensation models in Sections 2.1 and 2.2. Any centrally condensed model where a massive protostar originates from one condensation of roughly spherical shape will suffer a similar spacing problem. Two singular isothermal spheres with temperature 20 K, each truncated so its included mass could make a protostar of mass 10 $M_\odot$ with efficiency 0.5, would have a minimum center-to-center separation of 1.2 pc. To reduce their spacing to 0.1 pc would require kinetic temperatures exceeding 200 K, much higher than indicated by observations of $NH_3$ lines in cluster-forming regions (Jijina, Myers & Adams 1999).

Initial condensations for massive stars could have smaller diameters and closer spacings than in the models described here, if their clump densities were significantly increased. However the necessary increase would exceed available observational estimates. The mean column density of a spherical condensation of diameter $D$ which makes a protostar of mass $M$ with efficiency $\varepsilon$ is $1.4 \times 10^{25}$ cm$^{-2}$ or 53 g cm$^{-2}$ for $D = 0.01$ pc, $\varepsilon = 0.5$, and $M = 10\ M_\odot$. Such a high column density exceeds by a factor ~50 the column density 1 g cm$^{-2}$ seen in regions of massive star formation, which is also the value required by the turbulent core model of massive star formation (McKee & Tan 2003). This conclusion is independent of any particular model of condensation structure, such as the TNT model, and is independent of the premise of distributed infall durations.

A solution to this problem may lie in a highly nonspherical distribution of the initial mass. Condensations can be massive and closely spaced if they are highly elongated, so that their spacing is a small multiple of their shortest dimension, while most of their mass is extended along their longest dimension. Such a configuration can be filamentary, provided its filament or filaments have much greater aspect ratio than the filamentary condensations of Section 2.2, whose aspect ratio is unity.

Such elongated filaments can belong to networks with multiple nodes and branches, as seen in simulations of colliding flows (Banerjee et al 2010). If several such filaments converge on a central node, together they constitute a condensation of "stellate" shape which resembles the "hub-filament" configuration observed in numerous cluster-forming clouds (Myers 2009b). Such a structure also resembles the flow pattern of dense gas forming massive stars in some cluster-forming simulations (Smith, Longmore & Bonnell 2009, Smith et al 2010; Wang et al 2010).

The next section describes such massive star condensations which can have short nearest-neighbor spacings.

2.5. Hub-filament condensations

Numerical models of massive star formation by competitive accretion (e.g. Bate 2005) suggest that the initial configuration of mass which eventually accretes onto a massive protostar may not be bounded by a single closed surface, and may not have a simple shape. Even for a well-defined initial condensation, the accretion which provides most of the protostar mass may be driven more by global gravitational forces due to the clump which contains an initial condensation, than by local forces due only to the mass within the condensation. Nonetheless, it is useful to identify conditions where formation of a massive protostar can be represented by the isolated collapse of a single well-defined condensation, whose shape allows the close spacing of protostars in young clusters.

To overcome the spacing problem and to more nearly represent the filamentary nature of clump-fed accretion of massive protostars, a simple "hub-filament" condensation is described here. In this condensation the enclosed mass depends on spherical radius exactly as do the spherical and cylindrical models of Sections 2.1 and 2.2, over the range of radii $0 \le r \le R$.

This hub-filament condensation is a simple representation of a filamentary network having a single node at its center of gravity. In this model, the central hub has the same dependence of density on radius as in the core-clump model of Section 2.1, for radii $0 \le r \le r_0$. Here $r_0$ marks the transition from the spherically symmetric hub to multiple filaments whose axes radiate from the hub center. The filament axes are radial in 3D. Their azimuthal and polar angles are constrained to keep the system center of gravity at the hub center, so that the protostar forms there. The filament axes may be primarily coplanar and they need not have equal angular spacing. However they are not necessarily coplanar, in contrast to the hub-filament model of Myers (2009b). The transition radius $r_0$ is small enough to allow the spacing to neighboring condensations to match that seen in young clusters, i.e. $r_0 < 0.1$ pc. The density at $r_0$ is obtained from equation (1) and is denoted $n_0$.

To allow a simple example, for $r \geq r_0$ each filament is assumed to have uniform density $n_0$ while the density between filaments is assumed to be zero. In a more realistic model the filament density would decrease with increasing distance from the hub center and from the filament axis, and the filaments would have corelike condensations as in the filamentary chains of Section 2.2.

If $r_0$ is smaller than the radius $(A/B)^{3/4}$ where the core and spherical clump component densities are equal, the filament density $n_0$ has a high value characteristic of the thermally dominated core. This density is much greater than the mean density of the condensations in Sections 2.1 and 2.2. Such high filament density allows the volume filling factor of the hub-filament condensation to be much less than unity. As in the spherically symmetric model, most of the mass is in the clump gas, but here the clump consists of radially oriented filaments. Their low volume filling factor allows much closer spacing between neighboring hub-filament condensations than is possible for spherically symmetric condensations.

2.6. Volume filling factor

The fraction $F$ of the spherical volume within $r$ filled with gas available for star formation can be written as

$$F_s = 1 \qquad (13)$$

for $0 \leq r \leq R$ in the spherical condensation of Section 2.1, as

$$F_{hf} = 1 \qquad (14)$$

for $0 \leq r \leq r_0$ in the hub-filament condensation described above, and as

$$F_{hf} = u^{-3} + \frac{3A(1 - u^{-1})}{n_0 r^2} + \frac{9B(1 - u^{-7/3})}{7 n_0 r^{2/3}} \qquad (15)$$

for $r_0 \leq r \leq R$ in the hub-filament condensation, based on equations (2) - (5), where the dimensionless radius $u$ is defined by

$$u \equiv \frac{r}{r_0} \ . \qquad (16)$$

These volume filling factors are shown in Figure 3 for the parameters described above, $T = 15$ K and $\overline{N} = 1 \times 10^{22}$ cm$^{-2}$, and with $r_0 = 0.05$ pc and $R = 0.3$ pc. The corresponding masses within $r_0$ and $R$ are 2.1 and 63 $M_\odot$ respectively, and the uniform filament density is $n_0 = 3.5 \times 10^4$ cm$^{-3}$.

Figure 3 shows that star-forming gas in the hub-filament initial condensation takes up only 0.26 of the volume occupied by the spherically symmetric initial configuration within the same radius $R = 0.3$ pc, since the hub-filament gas has higher mean density than the spherically symmetric condensation gas. This difference allows the spacing of such hub-filament condensations to be limited by the transition radius $r_0 = 0.05$ pc rather than by the outer radius $R$

= 0.3 pc. This difference is possible because the hub-filament condensation is essentially a dense core of stellate shape, where dense gas is distributed in extended filaments around a central hub.

Some characteristics of such a dense filament are seen in $NH_3$ line observations of the Ophiuchus B region. The Oph B filament extends for about 0.3 pc, has greater mean density than the Oph A and Oph C condensations observed in the same way, and has relatively little differentiation between "core" and "intercore" gas, in contrast to most regions of isolated star formation (Friesen et al 2009).

The relation between dense gas volume filling factor and radius in equation (15) has various geometrical realizations, as discussed in the next section. These include "flaring" configurations where the filament number is fixed while the filament diameter increases with distance from the hub, and "branching" configurations where the filament diameter is fixed while the number of filaments increases with distance from the hub.

2.7. Filament configuration

The condensation structure can be quantified by considering a spherical shell between $r$ and $r + dr$, where the radius $r$ exceeds the hub-filament transition radius $r_0$. In this shell, the mass of dense filament gas matches the mass of the spherically symmetric condensation of Section 2.1, provided the gas with density $n_0$ occupies a fraction $f$ of the shell volume given by

$$f \equiv \frac{n_s}{n_0} \qquad (17)$$

where $n_s$ is the spherically symmetric condensation density given in equation (1).

If this dense gas is due to $N_{fil}$ identical filaments having radially directed axes and circular cross sections, the shell volume filled by dense gas is

$$4\pi r^2 f dr = 2\pi r^2 dr N_{fil}(1 - \cos\theta_{max}) \qquad (18)$$

where $\theta_{max}$ is the half the maximum angular width of each filament at radius $r$. Solving equation (18) for $\theta_{max}$ gives this angle as

$$\theta_{max} = Arc\cos\left(1 - \frac{2f}{N_{fil}}\right) . \qquad (19)$$

The distance from the hub center to a point on a filament axis can be written $x = r\cos\theta_{max}$. After substitution from equation (19), this distance is

$$x = r\left(1 - \frac{2f}{N_{fil}}\right) . \qquad (20)$$

Similarly, the distance from this point to the filament boundary, along a direction normal to the axis, can be written $y_{max} = r\sin\theta_{max}$ or

$$y_{max} = 2r \left[ \frac{f}{N_{fil}} \left( 1 - \frac{f}{N_{fil}} \right) \right]^{1/2} \qquad . \qquad (21)$$

Solving equation (21) for $N_{fil}$ gives

$$N_{fil} = \frac{2f}{1 - \sqrt{1 - (y_{max}/r)^2}} \qquad . \qquad (22)$$

For a fixed number of filaments $N_{fil}$, equation (21) indicates that the filament width $2y_{max}$ increases with increasing distance $r$ from the hub. For $r$ large compared to the hub radius $r_0$, $y_{max} \sim r^{2/3}$.

This "flaring" property derives physically from the assumption in Section 2 that the mass of the hub-filament condensation increases faster with radius than for a purely thermal core. This increase allows the spherically symmetric TNT model to produce protostars which match the IMF for cold spherical infall with equally likely stopping (Paper 1). In contrast, if the initial condensations were purely thermal cores, their analogous hub-filament condensations would have filaments with no flaring, and the mass function resulting from their cold spherical collapses with equally likely stopping would have no high-mass tail.

The filaments in nearby star-forming regions and in simulations of cluster-forming regions do not appear to flare outward from central locations as expected from equation (21). Observed filaments tend to have more nearly constant diameter, or to taper as they radiate away from a central hub, as seen in submillimeter images of the Orion Molecular Cloud (Johnstone & Bally 1999) or images of $N_2H^+$ 1-0 line emission from the young cluster in NGC 1333 (Walsh et al 2006). In some cases the filaments break up into finer radiating branches, as in the filaments extending NE from the L1688 complex in Ophiuchus (Myers 2009b). In simulations of cluster-forming clouds, central filaments also do not flare outward. Instead they are broader than their surrounding filaments, and they tend to fork and branch outward as part of a filamentary network (Banerjee et al 2009, Wang et al 2010).

Such outward branching of filaments may also be expected in the present hub-filament model, for filaments whose radius $y_{max}$ is constant. Then equation (22) indicates that the number of filaments must increase with increasing distance $r$ from the hub. If $y_{max}$ is comparable to the hub radius $r_0$, the number of filaments for $r >> r_0$ increases as $N_{fil} \sim r^{4/3}$.

This increase in the number of filaments with radius cannot be realized in a single condensation if all its filamentary extensions are exactly radial, since then branching occurs only at the origin, in the hub center. However this increase can be approximated if a small number of "guide" filaments curve outward along flaring, nonradial paths. Then the remaining filaments can branch outward from the guide filaments, along radial lines.

Such a nearly radial configuration is illustrated in Figure 4, which shows a central hub radiating a coplanar system of branching filaments. For simplicity only one quadrant of the system is shown. There, a single radial filament branches into two guide filaments which flare outward. These guide filaments originate radial branches. The radial position $r$ where each new branch starts is set when equation (22) has $N_{fil} = 4i$. Here $i = 1, 2, 3...$ is an integer and the factor of four is due to the fourfold symmetry of the branching pattern. Thus as the radial distance from the hub increases, this quadrant first shows a single filament, then two, then three, increasing to seven filaments at the largest radius.

Figure 4 is drawn for parameter values $T = 15$ K, $\overline{N} = 1 \times 10^{22}$ cm$^{-2}$, $R = 0.3$ pc, and $r_0 = y_{max} = 0.05$ pc as in Section 2.6. The lines mark the filament axes. When the filament width is taken into account, neighboring filaments will blend as they converge. The pattern in Figure 4 resembles that seen in the Serpens South cluster, where the hub which harbors numerous protostars radiates five nearly radial filaments of extinction over $\sim 0.6$ pc (Gutermuth et al 2008).

The configuration in Figure 4 is an initial configuration, and its shape would change as it contracted under its self-gravity. Much of the flow can be expected to be radial, along the radial filaments. However these filaments would not maintain their initial constant density. Instead their density would increase inward, as their length scale shrinks and as they begin to merge with their neighbors. The evolution of such contracting, converging filaments can be seen in the colliding flow simulations of Vázquez-Semadeni et al (2007, Figure 4).

Hub-filament condensations have no mass associated with joining one condensation to its neighbor, unlike the linear filament condensations in Section 2.2. However, neighbor hub-filament condensations can alter the gravitational infall of an individual hub-filament condensation by adding to the mass enclosed within a given radius. In the example of Figures 3 and 4, the volume filling factor is greater than 0.5 for radii less than 0.14 pc, so a given hub of radius 0.05 pc can have at most one identical neighbor hub in a contiguous position. Such a condensation has a relative increase in mass enclosed by its outer radius, due to its close-packed neighbor, by a factor less than 2. If the infall of such a paired system results in a single protostar, its mass exceeds that of an otherwise similar isolated system by a factor less than 2, and its mean mass accretion rate would increase by a factor less than $2^{3/2} = 2.8$. If instead the infall produced an equal-mass binary, each of its member masses would be the same as for an isolated system, and each of them would have a mean mass accretion rate greater by a factor $2^{1/2} = 1.4$.

This example suggests that hub-filament condensations having the same individual profile of mass with radius as a spherical condensation would yield more massive stars and would form with greater mass accretion rates than their spherical counterparts, due to their increased opportunity for close packing. However in the foregoing example the relative increases in mass and mass accretion rate are of order a factor of two. In a mass function histogram with factor of two bins, the change from the spherical case would be difficult to notice. It would be useful to investigate this effect in more detail with numerical calculations.

The structure in Figure 4 is one of many possible initial branching configurations, whose number of filaments increases with radial distance as in equation (22). The similarity of Figure 4 to some observed clouds is suggestive. However, a more meaningful comparison with observations requires numerical studies which can follow the formation and evolution of such filamentary structures with careful variation of parameters.

These results suggest that the spacing of protostars in a young cluster constrains initial condensation models of relatively massive protostars. As the mass of the protostar increases, it becomes more difficult to represent its initial configuration as a single condensation with similar extent in each direction, such as the spherically symmetric core-clump condensation of Section 2.1 or the linear filament condensation having aspect ratio close to unity, as in Section 2.2. For protostars of mass exceeding a few solar masses, the required extent of such initial condensations is typically a few 0.1 pc, too great to match the spacing of protostars in dense clusters.

Instead it is necessary to invoke a condensation with low volume filling factor, where the hub or core contains a small fraction of the condensation mass, while most of the mass is in dense radiating filaments having high density and low volume filling factor. Such a stellate shape does not match the conventional picture of a condensation. Nonetheless it approximates the organization of dense gas in some cluster-forming regions, and in some simulations of clouds which eventually form clusters.

## 3. Discussion

This paper extends the result of Paper 1, which showed that a simple model of star formation in clusters can reproduce the IMF. In this model, each protostar originates in a single condensation whose profile of mass with radius increases approximately linearly with radius in a thermal "core" and more rapidly with radius, approximately as $r^{7/3}$, in the core environment or "clump." This mass profile is similar to those of the spherically symmetric models known as "thermal-nonthermal" models (TNT; Myers & Fuller 1992) and "two-component turbulent core" models (2CTC; McKee & Tan 2003). When an ensemble of such condensations undergoes cold spherical infall with equally likely stopping, the distribution of protostar masses matches the IMF.

This paper develops condensation models having the same dependence of enclosed mass on spherical radius, but whose density is spherically symmetric on small scales and filamentary on large scales. These models provide a much better match to the large-scale structure of gas and protostars in cluster-forming regions than do condensations which are purely spherically symmetric.

3.1. Filamentary condensation models

Analytic models which are spherically symmetric are convenient for describing condensations whose density structure is space-filling and which has similar extent in all three space dimensions. For many applications, modest departures from spherical asymmetry can be represented by modifications or extensions of a spherically symmetric model (e.g. Bertoldi & McKee 1992).

However, spherically symmetric models can be misleading when representing structures whose long and short dimensions are very different. This point has become more relevant as observations have revealed the large-scale structure of star-forming clouds to have an increasingly important filamentary character (Molinari et al 2010), including stellate condensations having radiating filaments (Myers 2009b). Similarly, numerical models of colliding flows and collapsing turbulent clumps show filamentary networks as a prominent feature of their structure (Banerjee et al 2009, Wang et al 2010).

It is therefore important to develop and apply appropriate analytic representations of filamentary structure in star-forming condensation models. The models presented here provide simple descriptions where one filament is associated with many cores (Sections 2.2- 2.3), and where one core is associated with many filaments (Sections 2.5-2.7). It remains to develop models which have a more explicit relation between their structure and the gravitational, magnetic and pressure forces which operate on them.

3.2. Spherical collapse of filamentary condensations

The filamentary models in Sections 2.2 and Sections 2.3-2.5 have the same dependence of mass on radius as the spherically symmetric models of Section 2.1, so they have the same duration of cold spherical infall as a function of radius, and as a function of free fall time. This similarity is important to the calculation of their protostar mass functions in Section 3. Cold spherical infall is an approximation to their true dynamical motions, which also depend on magnetic forces, on pressure forces, due to both thermal and turbulent processes, and on rotational motions, disk accretion, and nonsteady accretion associated with disk instabilities. For spherically symmetric condensations, this approximation is represented by the accretion efficiency $\varepsilon$ as in Paper 1. The protostar mass uncertainty due to this approximation is estimated to be a factor ~2, similar to the width of one mass bin in many observational determinations of cluster mass functions (e.g. Luhman et al 2003), and smaller by a factor ~ 9 than the FWHM of the IMF of Kroupa (2002).

For initially smooth filamentary condensations, an additional source of uncertainty in the relation of protostar mass to infall duration is the tendency for uniform bodies to first collapse toward the filamentary symmetry axis and then to fragment into smaller condensations (Lin, Mestel, & Shu 1965). However, this tendency depends on the initial shape of the filament, since SPH simulations indicate that prolate spheroids tend not to fragment near their ends, unlike truncated cylinders, but instead are likely to fragment only near their centers (Nelson & Papaloizou 1993).

For the linear filament condensation of Section 2.2-2.3, the strong central concentration of its spherically symmetric core is expected to dominate the early stages of collapse toward the center of gravity. Thus no further fragmentation is likely, and it is expected that the accretion efficiency for the infall of the spherically symmetric and linear filament condensations will not differ significantly compared to the factor of ~2 uncertainty in protostar mass described above. It would be useful to study the collapse of this condensation numerically to better understand its dependence of protostar mass on infall time.

For the hub-filament condensation of Section 2.5-2.7, the character of the infall may depend on the relative importance of global and local gravity, as discussed for the infall of a horizontally truncated sheet by Burkert & Hartmann (2006). If most of the mass is in a few widely spaced filaments, the filaments may have time to develop significant internal structure from gravitational fragmentation as they globally contract toward the hub. However if most of the mass is in a dense layer of filaments (Schmid-Burgk 1967, Myers 2009b), global gravity may have a shorter dynamical time, and so may be more dominant over local fragmentation.

The simulations of Wang et al (2010) show rapid infall of dense gas along branched filaments onto a central region, where massive stars form at early times in more central locations, followed by lower-mass stars, near the center and in the filaments. This study extends that of Banerjee et al (2009) to finer resolution. Its hydrodynamic simulation shows filaments with similar shape to those deduced here in Figure 3, and demonstrates that inflow along converging filaments can form massive stars, as also suggested in this paper.

However, in these simulations the rapid accumulation of dense gas is due in its early stages to large-scale turbulent compression, followed by more localized gravitational infall. Furthermore the filaments in these simulations are substantially larger and less dense than those considered in the present hub-filament model. As with the linear filament condensations, simulations with initial conditions more appropriate to the present model are needed to compare the mass functions resulting from the initial condensations proposed here.

The contrast between filamentary and spherical geometry also affects some mechanisms of dense gas dispersal, and therefore the mean infall stopping time may differ between the two configurations. When dispersal of a condensation is due to the winds and outflows from a central protostar, a spherical condensation will have no preferred direction of escape, whereas a filamentary condensation will offer the least resistance in the plane of the protostar normal to the filament axis. Thus spherical condensations may experience dispersal from internal protostars over a wider range of angles than for filamentary condensations. Such condensations will tend to retain their shielding from neighboring protostars in the same filament, resulting in a longer mean stopping time than for otherwise similar spherical condensations.

Similarly, a recent numerical study found that dense filaments tend not to erode significantly due to photoionization from nearby O stars in a young cluster (Dale & Bonnell 2011).

On the other hand, accretion may stop due to dynamical ejection from a multiple system, or to gravitational competition against neighboring masses, and these processes may be largely independent of internal condensation geometry. It seems likely that several mechanisms can contribute to the termination of accretion, as discussed in Paper 1, but their relative importance remains unclear. More detailed observational and numerical studies are needed to better understand the contributions to the termination of accretion, and how they differ from one condensation to the next.

3.3. Relation to other models

The models in this paper combine some elements of the "monolithic collapse" model for massive stars (McKee & Tan 2003) and for low-mass stars (Alves, Lombardi & Lada 2007), hereafter MC; and of the "competitive accretion" model (Bonnell et al 1997, Smith, Longmore & Bonnell 2009; hereafter CA) described in Section 1. A more detailed description of the MC and CA models is given in Peters et al (2010). For ease of comparison the present models are termed "randomly stopping accretion" or RSA.

RSA models resemble the MC picture in that protostars form by isolated gravitational collapse in condensations having a well-defined structure. This structure includes a core component and an environment component. Low-mass stars form primarily from the core component, forming in the same way in clusters as in regions of isolated star formation.

In contrast to star formation by MC, the distribution of RSA protostar masses arises mainly from the distribution of infall durations, not from a wide range of initial condensation masses. The initial condensation masses are considered similar in RSA models because the range of Jeans masses in star-forming gas is much smaller than the range of stellar masses in the IMF. The wide range of observed core masses, often used to justify MC models, is ascribed here to cores which have not completed their fragmentation, which are blended by projection and insufficient observational resolution, or which are likely to disperse before forming any stars. In this view, the similarity of observed distributions of core masses to the IMF is coincidental and not fundamental, as suggested by Hatchell & Fuller (2008).

RSA models resemble the CA picture in that the protostar mass is set mainly by the duration of infall, so the protostar mass is independent of the mass of the core where it forms. Also as in the CA picture, most of the mass of a massive star originates farther from its center than the core radius, in gas which flows toward the clump center of gravity along converging filaments. However, in the RSA model one initial condensation makes one protostar, unlike CA models where an initial clump makes many stars in more complex fashion. Further, the RSA model lacks the stellar dynamics of CA models, and relies only on stationary accretion rather than on both stationary and moving accretion.

The RSA model goes beyond the MC and CA pictures in that it explicitly specifies the distribution of infall durations, and thereby approximates the IMF with condensation models matching observed condensations in their temperature and mean column density. Also, describing initial condensations with both spherical and filamentary components represents the structure of young clusters more realistically than do the MC and CA pictures. It provides a stellate representation of condensations forming massive stars, allowing such protostar birthsites to have the short observed spacings of protostars in young clusters, and the filamentary channels of mass flow seen in some simulations of massive star formation.

3.4. Limitations

As the mass of the protostar increases, the models presented here are limited by the assumption that every protostar originates exclusively from a single initial condensation of simple shape. This assumption is basic to models of isolated low-mass star formation (e.g. Shu, Adams, & Lizano 1987), but for cluster protostars of increasing mass, this premise is challenged by the observation that cluster protostars have central spacings significantly less than 0.1 pc, while according to many condensation models, condensation diameters must significantly exceed 0.1 pc.

The solution proposed in Sections 2.5 - 2.7 is that each initial condensation has very different short and long dimensions, with a low volume filling factor. The resulting stellate condensation has a relatively small core and radiates numerous filamentary arms, which contain most of the condensation mass. This shape allows much closer condensation spacings than do more conventional condensations. It resembles hub-filament configurations seen on larger scales in observed cluster-forming regions and in simulations of colliding flows (Banerjee et al 2009) and collapsing turbulent clumps (Smith et al 2010, Wang et al 2010).

This solution may be a useful representation of a condensation which has enough mass to form a massive protostar. This center of a condensation with this shape can lie closer to the center of a neighboring star-forming condensation than is possible for condensations of simpler shape. As with the spherical and filamentary condensations considered earlier, the mass of its stellar product can be approximated by isolated gravitational collapse in the assumption of cold spherical infall with equally likely stopping.

The cluster spacing problem has the premise that the initial condensations for the cluster protostars are all present at the same initial moment. Since the star-forming life of the cluster is much longer than the free-fall time of its condensations, it seems likely that many generations of star formation contribute to the observed concentration of protostars. If so, the crowding problem might be alleviated by spreading the star formation out in time. Thus first-generation star formation could free up volume which is then available for later-generation condensations to form and to produce later-generation stars. However, this picture has seemingly restrictive requirements in both space and time: such new condensations must form in newly available locations and must form on the free-fall time scale of the previous condensations.

It is also possible that the paradigm of isolated star formation does not apply to the densest parts of young clusters. The idea that one condensation makes one protostar may not describe the multiple nature of clustered star formation, especially for more massive stars. Thus a complex of closely spaced stellate condensations might be indistinguishable, in both observations and simulations, from a centrally condensed clump forming multiple protostars along filamentary flow paths, as in the simulations noted above. Such a complex would be more difficult to parse into star-forming components than would a complex of components of simpler shape having the same spacings. If so, this confusion may limit of applicability of isolated models of star formation to clusters.

The calculations in this paper rely on gas temperature and column density estimates for regions forming clusters with ~ 100 members. This approach limits the scope of the model and does not allow for regions with far fewer members, such as groups with ~ 20 members (Kirk & Myers 2011), or super star clusters with thousands of members. It will be important to understand conditions under which the concepts described here can be extended to smaller and larger clusters.

3.5. Relation to observed young cluster structure

Among the three models considered here, the linear filament model provides a much closer approximation to the pc-scale lumpy filaments in and near young clusters than does the spherical condensation model. In addition to the examples cited in sections 1 and 2.3, the linear chain of three SiO condensations in the infrared dark cloud G035.39-00.33 (Jiménez-Serra et al 2010) and the linear chain of five 0.88 mm continuum condensations in the G028.34+0.06 cloud (Zhang et al 2009) closely resemble the equal spacing filamentary chain in Figure 1.

Some cluster-forming regions also appear to have stellate condensations, where multiple filaments of dense gas radiate from a single hub, on scales of 0.1 pc, as in the high-resolution observations of $N_2H^+$ 1-0 in NGC 1333 (Walsh et al 2006), and on pc scales, as seen in numerous nearby groups and clusters, and in more distant infrared dark clouds (Myers 2009b).

However, many cluster-forming regions appear too crowded and complex with currently available resolution and sensitivity to parse into distinct stellate structures. High-resolution, wide-field interferometric observations of young clusters, with currently operating imaging arrays, and with the forthcoming ALMA, can be expected to test the relevance of the stellate condensation model. Such observations are also needed to discriminate between a model of distinct "isolated" stellate condensations and a more cooperative star-forming model where a filamentary clump makes multiple stars, as suggested by some simulations.

## 4. Conclusions

This paper presents three condensation models which may describe initial conditions for star formation in clusters. A "spherical" condensation has a spherical core in a spherical clump. It consists of a small-scale thermal core and a surrounding clump with shallower density gradient, as in TNT models (Myers & Fuller 1992) or 2CTC models (McKee & Tan 2003). A "linear filament" condensation has a spherical core in a cylindrical clump. The aspect ratio of this clump is unity, and these filamentary condensations are organized into linear filamentary chains. A "hub-filament" condensation has stellate shape, with a spherical core radiating multiple dense filaments.

Each of these geometrically distinct models has the identical distribution of mass with enclosed spherical radius, needed to generate the IMF for cold spherical infall with equally likely stopping. Under more realistic conditions it is expected that the mass functions resulting from collapse of these three condensations will be similar for low-mass stars and will differ more for massive stars.

The standard set of parameters providing good match to the IMF over the protostar mass range 0.01 - 10 $M_\odot$ is kinetic temperature $T = 15$ K, maximum radius $R = 0.15$ pc, mean column density within $R$ $\bar{N} = 10^{22}$ cm$^{-2}$, accretion efficiency $\varepsilon = 0.5$, and mean infall duration $\bar{t}_f = 0.05$ Myr.

The main conclusions are:

1. Spherically symmetric condensations do not match the large-scale filamentary appearance of star-forming gas.

2. The linear filament model provides a more realistic representation of parsec-scale filaments of cluster-forming gas, and of their embedded cores, than does the spherical model. When such filaments converge toward a cluster center, they concentrate the birthsites of low-mass stars. A system of ten radial filaments having radius 0.4 pc and condensations 0.2 pc apart provides a mean surface density of birthsites ~40 pc$^{-2}$, in good agreement with observed protostar surface densities in young clusters.

3. Initial condensations having simple shapes and aspect ratios near unity cannot pack closely enough for the spacing of protostar birthsites to match that of protostars in young cluster centers, typically a few 0.01 pc. To make a protostar of 10 $M_\odot$ with efficiency 0.5 from a spherically symmetric condensation of diameter 0.01 pc requires initial mean column density exceeding $10^{25}$ cm$^{-2}$, or mass surface density exceeding 50 g cm$^{-2}$, independent of any model of condensation density or infall stopping. These column densities greatly exceed observed values.

4. The foregoing spacing problem for massive protostars appears to exclude spherically symmetric infall and instead appears to require infall along paths which have a low volume filling factor.

5. Stellate condensations having a central core and multiple radiating filaments can provide enough mass and sufficently low volume filling factor if their filaments are sufficiently numerous, long and dense. Most the mass of such a hub-filament condensation is in its filaments, which may flare or branch outward with increasing distance from the hub. This structure resembles that seen on scales of 0.1 - 1 pc in cluster-forming regions and on pc scales in some simulations of colliding flows and collapsing clumps.

6. Models where the mass of each protostar comes from one well-defined condensation become more complex and less useful, as protostar mass increases and as the spacing of protostar birthsites decreases in a young cluster. Stellate condensations of low volume filling factor can match observational requirements on protostar mass and spacing, but a close-packed configuration of such condensations may be indistinguishable from a cluster-forming clump forming multiple protostars along filamentary flow paths.

**Acknowledgements** This paper has benefited from helpful discussions with Fred Adams, Lori Allen, Matthew Bate, Paola Caselli, Rob Gutermuth, Lee Hartmann, Helen Kirk, Charlie Lada, Tom Megeath, Ralph Pudritz, Rowan Smith, Jonathan Tan, and Qizhou Zhang. The anonymous referee made several constructive and useful comments and suggestions. Many discussions useful for this work occurred at the conference "The Origins of Stellar Masses" in Tenerife, Spain, in October 2010. The author is also grateful for support and encouragement from Irwin Shapiro and Terry Marshall.

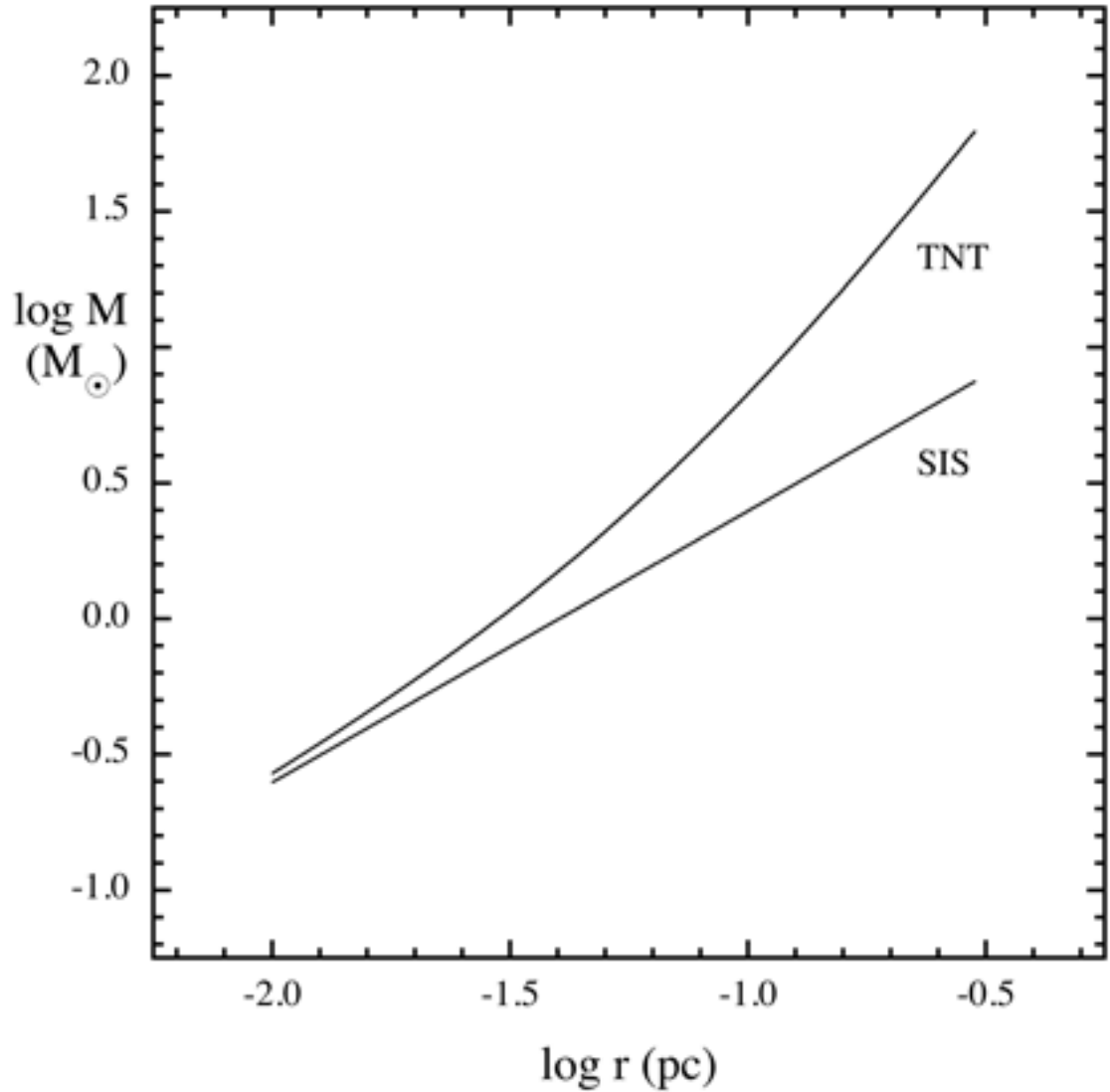

**Figure 1.** Mass as a function of radius for the spherically symmetric thermal-nonthermal (TNT) condensation described by equation (5), with core temperature $T = 15$ K, mean clump column density $= 10^{22}$ cm$^{-2}$, and maximum radius $R = 0.5$ pc. The mass profile resembles that of a singular isothermal sphere (SIS) at 15 K at small scales, but increases more rapidly at large scales, where the clump component dominates the mass.

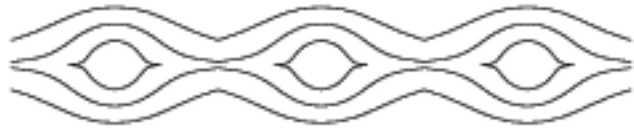

**Figure 2.** Contours of column density 15, 20, and 30 × $10^{21}$ cm$^{-2}$ for a chain of three linear filament condensations, for core temperature 20 K, central clump column density 20 × $10^{21}$ cm$^{-2}$, and condensation spacing 0.2 pc. This structure resembles filaments with embedded cores in cluster-forming regions. Each condensation has the same dependence of mass on radius as does a spherical TNT model.

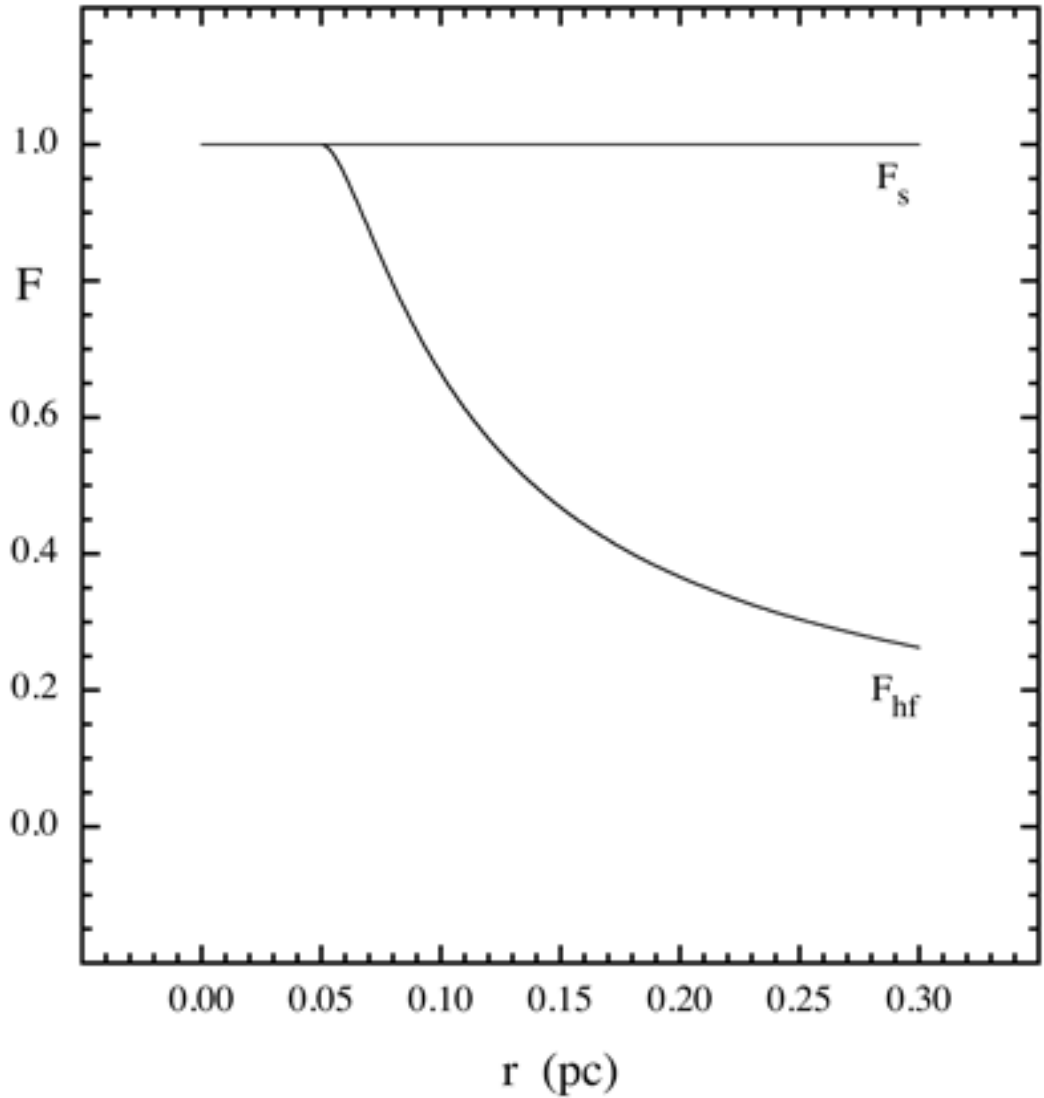

**Figure 3.** Volume filling factors of gas available for star formation, for two condensations having the same mass within $r$, for $0 < r < R = 0.30$ pc. The spherical condensation described in Section 2.1 has filling factor $F_s = 1$, while the hub-filament model has filling factor $F_{hf}$ declining from 1 to 0.26 as the radius increases from the transition radius $r_0 = 0.05$ pc to the maximum radius $R = 0.30$ pc. As radius $r$ increases, the filaments occupy an increasing fraction of the mass, but a decreasing fraction of the spherical volume within $r$.

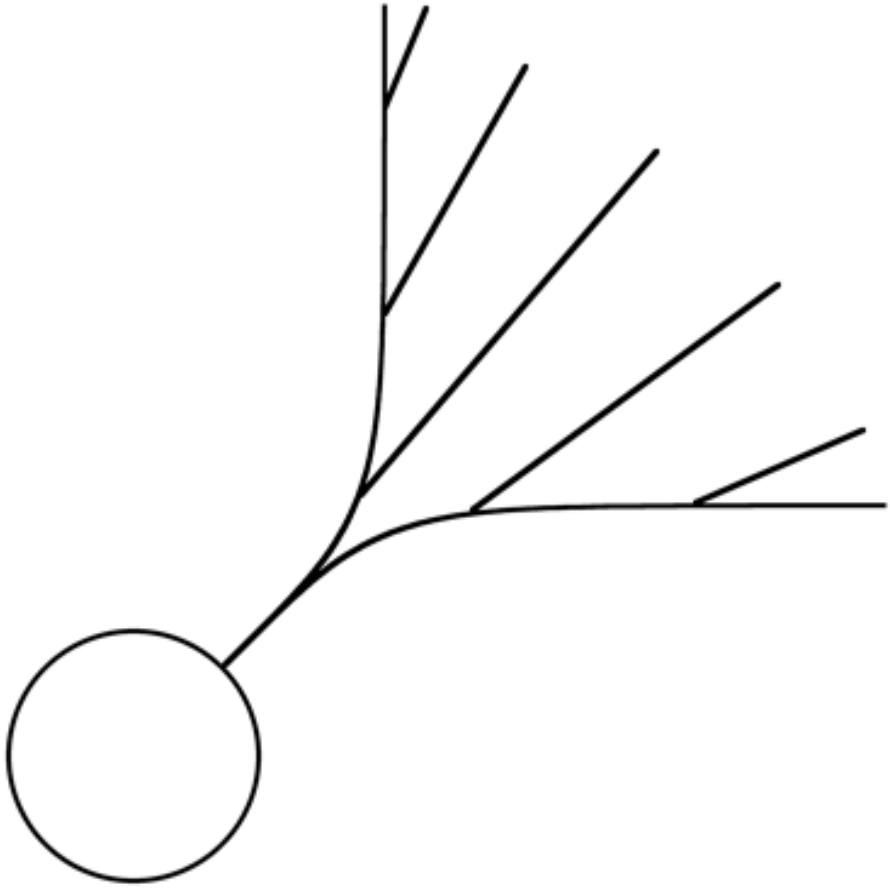

**Figure 4**. Schematic structure of a stellate hub-filament initial condensation with branching filaments of constant density, for the same condensation parameters as in Figure 3. This system has essentially the same mass-radius profile as do the spherical and linear filament condensations described in Sections 2.1 and 2.2. Its number of filaments increases with radius following equation (22). Its stellate structure allows shorter spacing between neighboring condensations than would be possible for the foregoing spherical and filamentary condensations. The circle represents the transition radius $r_0 = 0.05$ pc between the central hub and its radiating filaments. The lines represent the axes of coplanar filamentary branches in one quadrant of the structure.